# Astro2020 Science White Paper
## Technosignatures in the Thermal Infrared

**Primary thematic area: Planetary systems**, especially exobiology and the search for life beyond the Solar System

**Secondary thematic areas: Star and planet formation**, especially protostellar and protoplanetary disks and dust, **Resolved stellar populations and their environments**, especially stellar populations and evolution, **Galaxy evolution**, especially star formation rates and dust

**Principal Author: Jason T. Wright**
astrowright@gmail.com
525 Davey Laboratory
Department of Astronomy & Astrophysics and Center for Habitable Worlds
Penn State University
University Park, PA 16802
(814) 863-8470

**Co-authors**: **Erik Zackrisson,** Uppsala U., **Casey Lisse**, Johns Hopkins U.

**Abstract:**
*WISE, Gaia,* and *JWST* provide an opportunity to compute the first robust upper limits on the energy supplies of extraterrestrial civilizations, both for stars in the Galaxy (Kardashev Type II civilizations) and for other galaxies (Kardashev Type III civilizations). Together, they allow for a nearly-complete catalog of nearby stars with infrared excesses, which is valuable for both stellar astrophysics and searches for technosignatures; and of the MIR luminosities of galaxies, important for studies of galaxies' star and star-formation properties, but also for the identification of potential galaxies endemic with alien technology. *JWST* will provide the crucial mid-infrared spectroscopy necessary to identify the origin of these infrared excesses, advancing both traditional astronomy and searches for technosignatures. Such signatures are distinguished from dust by their lack of far-infrared emission and lack of association with star formation.

# The Rationale for Infrared Technosignatures

One of the most general technosignatures is the "waste" heat of industry; that is, the inevitable emission of high entropy energy after that energy has been used for some purpose. This method of identifying extraterrestrial technology is as old as radio SETI (Dyson 1960) but has hardly been pursued, at first because infrared astronomy was not sufficiently mature for such a search to be feasible until the 1980's, then later because of the general funding landscape for searches for extraterrestrial technology.

Searching for the thermal emission of technological energy use satisfies many of the desiderata of a good SETI search: the signature is as long-lived as the underlying technology, it is an inevitable consequence of all energetic technology use (indeed, required by the laws of thermodynamics[1]), and it is detectable not only with current astronomical instrumentation, but indeed much of the data needed to perform the search has been or will be taken for other astronomical purposes.

The primary drawback of the method is that, unlike communicative transmissions, infrared excesses of astronomical sources are common for natural reasons. This, however, also means **that searches for anomalous infrared signatures from nearby stars and galaxies will have good synergies with other parts of astronomy**, while still providing other forms of SETI with enriched target lists for more unambiguous technosignatures. Indeed, **the anomalies discovered in searches for waste heat will be inherently astrophysically interesting regardless of their nature**, and so satisfy Freeman Dyson's "First Law of SETI Investigations": "every search for alien civilizations should be planned to give interesting results even when no aliens are discovered."

---

[1] The only way to avoid excess waste heat on very long time scales (i.e. that which would not have been generated otherwise by, e.g., dust or a planetary surface) is to emit it at low entropy (for instance with powerful transmissions) but even this activity is subject to Carnot limits. It is sometimes incorrectly argued that any "advanced" alien technology would be too efficient to emit waste heat, but this confuses colloquial and physics definitions of "efficiency." For instance, computers on Earth today use several orders of magnitude fewer ergs per computation than those of the 1960's; but the *total* power consumption of computers today is several orders of magnitude more than in the 1960's, while the fraction of that power radiated away as "waste" heat remains 100%.

# Towards Robust Upper Limits of Energy Use

Interpreting infrared measurements of stars and galaxies serves searches for technosignatures in two ways: it identifies anomalous infrared sources for further study as both natural or artificial objects of interest (Djorgovski 2000), and it allows for upper limits to be computed (e.g. Wright, Kanodia, & Lubar 2018).

Because energy use is a general technosignature, it can be parameterized rather simply, for instance using the *AGENT* scheme of Wright et al. (2014) where, $\alpha$ is the fraction of starlight blocked by technology (for instance by solar panels or radiators, but the scheme is completely general) and $\gamma$ is the fraction of starlight re-radiated as heat at some characteristic temperature $T_{\text{waste}}$. In the optical, where megastructures might be detected in transit, the average decrement in flux over all directions is roughly $-\alpha$. In the thermal infrared, however, the flux boost is given by

$$\frac{\Delta F}{F} = \gamma \left(\frac{T_*}{T_{\text{waste}}}\right)^3 - \alpha$$

Since for starlight-fed industry, we expect $\gamma \approx \alpha$, this shows the brightening of sources in the infrared is a significantly stronger signal than the optical decrement. It is also more omnidirectional and persistent. **Together, these properties illustrate the power of the thermal infrared for producing sensitive upper limits on alien industry.**

This parameterization applies equally well to planets, stars, and galaxies (i.e. civilizations approaching Types I, II and III on the Kardashev (1964) scale). At the smallest scale, the infrared emission from a planet might reveal regions of industry on its surface (Kuhn & Berdyugina 2015). Observations of the infrared excess of a single star sets limits on circumstellar technology, and the integrated infrared light of a galaxy puts limits on pan-galactic energy use, for instance from a spacefaring civilization that has settled the galaxy.

# Necessary Advances to Realize the Thermal Technosignature Opportunity

Much or most of the necessary advances needed to realize this opportunity are already being made in the search for the most common natural confounders for waste heat searches.

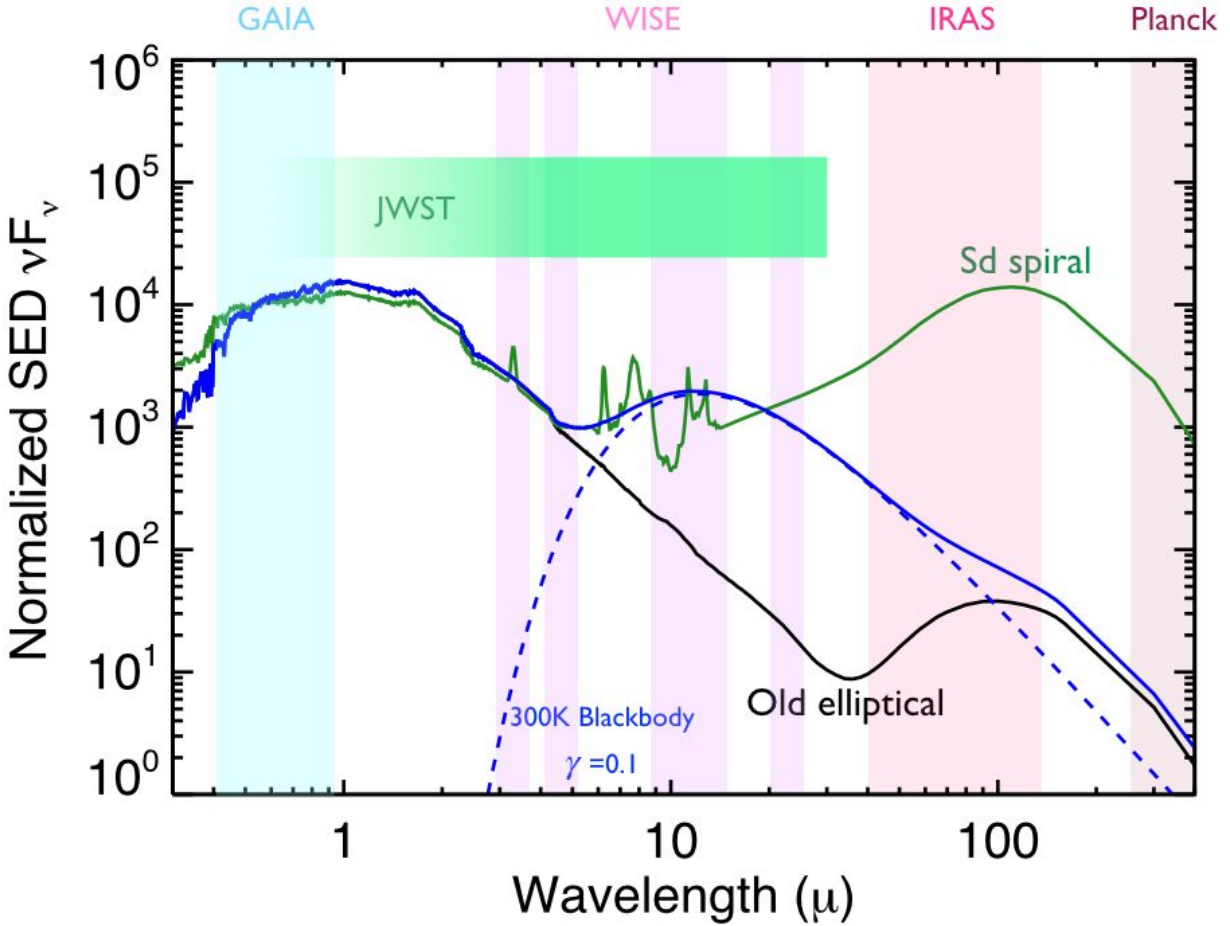

*Figure 1. SEDs from Silva et al. (1998) of an old elliptical galaxy (black) and a typical Sd spiral (green) normalized by their optical flux. In blue is the old elliptical with 10% of the starlight reprocessed as 300K waste heat (these qualitative results are insensitive to the precise choice of this temperature). Also shown are the wavelengths of 4 relevant all sky surveys as vertical stripes, and JWST's wavelength coverage as a horizontal stripe.*

Figure 1 compares the SEDs of a relatively dust-free elliptical galaxy, an Sd spiral, and the expectation for a dust-free old stellar population with 10% of the starlight reprocessed into 300K waste heat (blue). The optical and near infrared-colors of a dusty stellar population and a dust-free population with waste heat are similar, so *WISE* alone cannot distinguish these cases in integrated light (i.e. unresolved galaxies).

However, this degeneracy can be broken using infrared spectra to identify the PAH features expected from dust, or with far-infrared measurements. In particular, large amounts of waste heat at hundreds of Kelvins should generate large MIR-FIR colors that is not expected from dust, which usually has a cool component. Circumstellar waste heat shares many optical to MIR broadband characteristics with circumstellar dust, and so searches for and studies of debris disks can be easily and slightly modified to become searches for waste heat (for instance, the Disk Detective Zooniverse project could easily be repurposed for this).

Circumstellar dust can be distinguished from partial Dyson spheres by their spectra, high resolution imagery, and the fact that debris disks are rare among old stars. Indeed, debris disks created by second generation release of dust and gas from young planetesimals (asteroids, comets, KBOs) are often found around 10-1000 Myr old stars making their smaller, rockier and icier planets, but are almost never found around multiGyr old stars (Wyatt et al 2007, Wyatt 2008) as these systems become cleaned out via radiation pressure, Poynting-Robertson drag, collisions, and gravitational ejections in their maturity. The infrared excesses of the old stars tau Ceti and HD 68380 are good examples of the sorts of signals that are either unusually old debris disks (and, so interesting in their own rights) or evidence for extraterrestrial technology around nearby, sun-like, planet-bearing stars. (See white papers by Mennesson, Debes, and Su).

In other galaxies, waste heat must be treated statistically across the stellar population, and dust and waste heat can be distinguished angularly as well as spectrally. Resolved galaxies are amenable to approaches similar to Conroy & van Dokkum (2016), who have shown that the SEDs of individual pixels of resolved galaxies can be used to deduce the star-formation history and other properties of the underlying stellar populations. Since dust in galaxies is clumpy while the stars should be relatively well mixed due to shear and thermal velocities, **stellar population synthesis models that are modified to include a smoothly-varying waste heat component can be used to separate the dust and waste heat components of infrared excesses in galaxies.** For instance, waste heat should be present and conspicuous between the spiral arms of galaxies and throughout ellipticals, while dust should trace star formation and, at a lower level, the AGB stars. More detailed models of waste heat can produce correspondingly more detailed predictions for observable anomalies (Lacki 2018).

# The Opportunity in the 2020's

**The completion of the primary *WISE* (Wright et al. 2010) all-sky survey and release of the ALLWISE catalog have revolutionized the search for thermal technosignatures.** Its unprecedented, all-sky point-source sensitivity at 20μ provides the first opportunity for stringent measurements of $\gamma$ for most stars and galaxies.

For instance, Griffith et al. (2015) used the *WISE* colors of 100,000 resolved galaxies to identify those with anomalously large MIR-NIR colors (see Figure 2). They put the first (weak) upper limits on γ among these sources, with the principle confounding source being starburst galaxies. More detailed SED modeling over a wider range of wavelengths will make it possible to extend this work to unresolved sources.

The ALLWISE catalog has hundreds of millions point sources, most of them distant galaxies, which complicates searches for stars with anomalous levels of infrared emission. Fortunately, *Gaia* provides parallaxes for most of the nearby stars represented in *WISE*, providing a clean way to distinguish the Galactic sources in ALLWISE from the extragalactic sources. In total, tens of millions of stars with non-zero parallaxes from *Gaia* DR2 have ALLWISE detections that would allow their MIR excess to be constrained. **Together, *Gaia* and *WISE* provide an opportunity for the first strong upper limits on waste heat from stars in the Milky Way.**

These upper limits will be set by the most anomalous (but presumably natural) sources. **Follow-up of these sources to determine their nature will be best performed with mid-infrared spectroscopy, especially from *JWST*.**

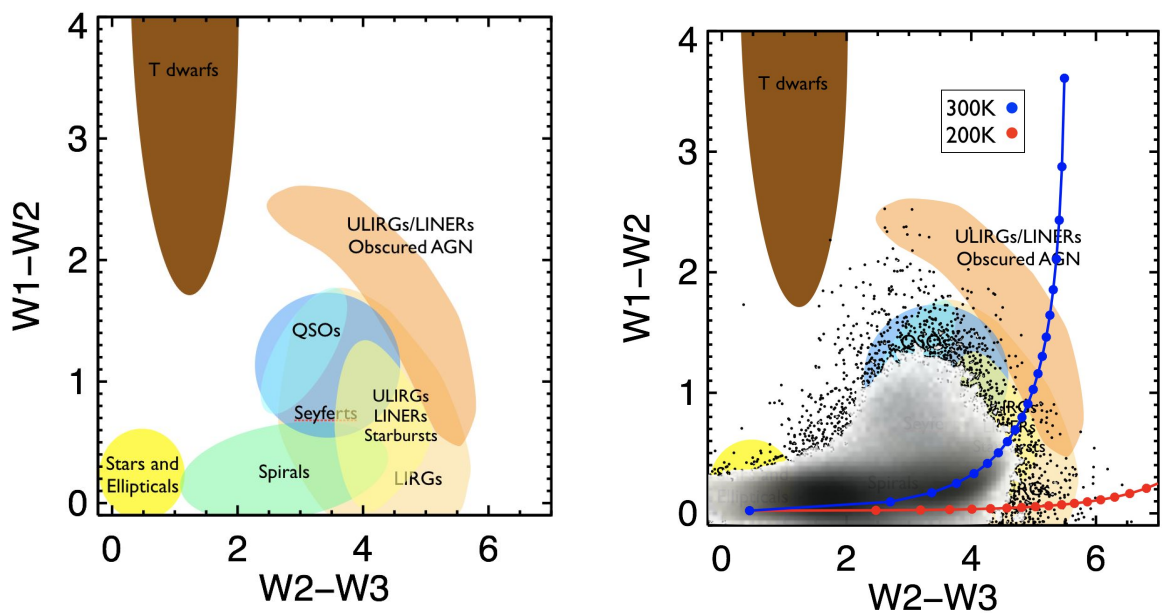

*Figure 2. Left: WISE colors of classes of natural sources, after Lake et al. (2012) and Jarrett et al. (2011). Right: Sources common to WISE and the 2MASS XSC as black points and logarithmic grayscale where point density is high. Also shown are the loci of stars or elliptical galaxies with starlight reprocessed at 200K (red) and 300K (blue) from γ=0 (no waste heat, lower left) to γ=1 (100% reradiation of starlight, upper right in blue, off figure to right in red) in increments of 0.05. After Wright et al. (2014).*